\documentclass[twocolumn,amsmath,amssymb,aps,prl]{revtex4-1}
\usepackage[latin9]{inputenc}
\usepackage{amsmath}
\usepackage{amssymb}
\usepackage{graphicx}
\usepackage{esint}

\makeatletter
\usepackage{dcolumn}% Align table columns on decimal point
\usepackage{array}
\usepackage{bm}
\usepackage{txfonts}
\usepackage{wasysym}
\usepackage{subfigure}
\usepackage{color}
\usepackage{array}
\usepackage{epsfig}
\usepackage[colorinlistoftodos]{todonotes}

\makeatother

\begin{document}

\title{Interpreting machine learning of topological quantum phase transitions}

\author{Yi Zhang$^{1,2}$}
\email{frankzhangyi@gmail.com}
\author{Paul Ginsparg$^1$}
\email{ginsparg@cornell.edu}
\author{Eun-Ah Kim$^1$}
\email{eun-ah.kim@cornell.edu}

\affiliation{$^{1}$Department of Physics, Cornell University, Ithaca, New York 14853, USA}
\affiliation{$^{2}$International Center for Quantum Materials, Peking University, Beijing, 100871, China}

\date{\today}

\begin{abstract}
There has been growing excitement over the possibility of employing artificial neural networks (ANNs) to gain new theoretical insight into the physics of quantum many-body problems. ``Interpretability'' remains a concern: can we understand the basis for the ANN's decision-making criteria in order to inform our theoretical understanding? ``Interpretable'' machine learning in quantum matter has to date been restricted to linear models, such as support vector machines, due to the greater difficulty of interpreting non-linear ANNs. Here we consider topological quantum phase transitions in models of Chern insulator, $\mathbb{Z}_2$ topological insulator, and $\mathbb{Z}_2$ quantum spin liquid, each using a shallow fully connected feed-forward ANN. The use of quantum loop topography, a ``domain knowledge''-guided approach to feature selection, facilitates the construction of faithful phase diagrams and semi-quantitative interpretation of the criteria in certain cases. To identify the topological phases, the ANNs learn physically meaningful features, such as topological invariants and deconfinement of loops. The interpretability in these cases suggests hope for theoretical progress based on future uses of ANN-based machine learning on quantum many-body problems.
\end{abstract}

\maketitle

\section{Introduction}

There has been much recent activity in the quantum matter community applying ANN-based machine learning to synthetic \cite{Carleo2016,  Deng2017, GaoXun2017, Junwei2017, Torlai2018, Glasser2018, Choo2018, Melko20161, Kelvin2016, Simon2016, qlt2016, FrankMLZ2, YiTing2018, Zhaihui2018L, Zhaihui2018B, Greplova2019, Ohtsuki2016, Olek2020, Tsai2019, Caio2019, Nieva2019, Schwab2014, Seif2019} and experimental \cite{mlstm2018, demler2018ml, Weitenberg2018ml, Jeremy2019} quantum matter data. These efforts exploit the ability of ANN models to provide effective approximations of functions in high-dimensional spaces. ANNs can thus represent either many-body wave functions or complex mappings between many-body Hilbert space and associated emergent properties. Indeed ANN-based variational studies \cite{Carleo2016,  Deng2017, GaoXun2017, Junwei2017, Torlai2018, Glasser2018, Choo2018} and phase detection \cite{Melko20161, Kelvin2016, Simon2016, qlt2016, FrankMLZ2, YiTing2018, Zhaihui2018L, Zhaihui2018B, Greplova2019, Ohtsuki2016, Olek2020, Tsai2019, Caio2019, Nieva2019} have successfully reproduced known results, even for difficult cases such as topological characters and many-body localization. Even the simplest ANN with a single hidden layer can approximate any target function with a sufficiently wide hidden layer \cite{Cybenko1989}; though in practice, such an architecture would be difficult to train efficiently, so multi-layer architectures are preferred. Insofar as the goal of theoretical quantum matter physics is some fundamental understanding of its underlying properties, however, it would be disappointing if the ANN operated as a black box, giving little insight into the basis for its predictions or results. More generally, ``interpretability'' is an important challenge in many areas of algorithmic Artificial Intelligence, and is of particular importance in the use of ANNs as tools for scientific research.

The more expressive is a machine learning model, the harder it can be to interpret. It is difficult to characterize a function on a high-dimensional domain if it has an enormous number of parameters, and no obvious symmetries to permit easy visualization. If an ANN trained on a large number of labeled samples is able to predict with high precision the expected properties of new sample data, then we have certainly made progress. But if we are unable to extract from the ANN the specific features of the data, and combinations thereof, that it uses to make those predictions, then we have not achieved a deeper insight into the physics of the system. Our goal in physics is to develop some more compact formulation of the crucial degrees of freedom, derive from that some more general intuition into the system's behavior, and use that to develop analytic methods or physical laws that can be extended to a wide variety of related phenomena.  A black box predictor that works only on a specific class of examples, and gives no insight into how it makes predictions, would be fundamentally unsatisfying as a tool for theoretical physics.

To date, there have been several preliminary endeavors on ``interpretable'' machine learning in problems of quantum matter. For instance, support vector machines
%To date, ``interpretable'' machine learning in problems of quantum matter has been restricted to support vector machines (SVMs). Specifically, SVMs
have detected features of the order parameters in various spin models \cite{MelkoIntp2017, LiuKe2018, LiuKe2019}, the Hamiltonian constraints in gauge theories \cite{MelkoIntp2017}, and the level statistics in the many-body localization transition \cite{LeiWang2019SVM}.
The support vector machines are intrinsically linear classifiers, based on finding a hyperplane to separate an already curated feature set, hence more easily interpreted. The more generally adopted ANN, on the other hand, can take an unmanageably large set of raw features, and transform and combine them in arbitrarily complex ways by way of millions or billions of learned parameter values. Its predictions are opaque unless we can determine how it has rearranged, amplified, and combined them into effective degrees of freedom that govern the phenomena of interest. One viable interpretive method is Taylor expansion, which has succeeded in the extraction of certain order parameter expressions \cite{Sebastian2017} and even laws of physics \cite{WangCe2019, Yadong2018}; however, its complexity quickly evolves out of control as the ANN architecture or the target function becomes more complicated.

Here we consider three investigations of topological quantum phase transitions, each using a shallow fully connected feed-forward neural network. The quantum phase transitions are between topologically trivial states and three distinct topological phases in two spatial dimensions: a time-reversal symmetry breaking Chern insulator (CI), a time-reversal invariant $\mathbb{Z}_2$ topological insulator (TI) and a $\mathbb{Z}_2$ quantum spin liquid (QSL). Of the three cases, the ANN-based phase diagram for a Chern insulator \cite{qlt2016} and $\mathbb{Z}_2$ quantum spin liquid \cite{FrankMLZ2} have been previously obtained by two of us. The case of the ANN-based phase diagram for the $\mathbb{Z}_2$ TI, as far as we know, is first obtained in section~\ref{sec:QSH} here, although the model of \textcite{QSHE2005} is well-known. In all three cases, the topological order is detected using only a simple shallow ANN, by using the physically motivated features introduced in Ref.\cite{qlt2016}, designated quantum loop topography (QLT). The QLT consists of a semi-local gauge invariant product of two-point functions from (variational) Monte Carlo instances. The specific geometry of the QLT is guided by characteristics of the phase itself, i.e., it is based on ``domain knowledge''. Given the simplicity of our ANN, and its ability to interpolate between QLT and the topological phases of interest, it is plausible that insight into this physics can be derived by probing the ``interior" of the ANN to illuminate properties of the function it has learned.

In this paper, we probe trained ANNs that yield correct topological quantum phase diagrams for the three cases of interest. We find interpretations of the ``learning'' of these ANNs to fall in two classes: (1) a linear function corresponding to a topological invariant, and (2) non-linear functions that build non-local and non-linear observables from our QLT inputs. The CI case in section \ref{sec:CI} falls into class (1), and the $\mathbb{Z}_2$ TI and the $\mathbb{Z}_2$ QSL cases fall into class (2). The remainder of this paper is organized as follows. In section \ref{sec:CI}, we review the QLT and ANN-based phase detection for the Chern insulators, and interpret the trained ANN. In section~\ref{sec:QSH}, we obtain an ANN-based phase diagram for the $\mathbb{Z}_2$ TI using a new QLT, and again interpret what the trained ANN has learned about the system. In section~\ref{sec:QSL}, we examine an ANN trained to detect the $\mathbb{Z}_2$ quantum spin liquid phase, and interpret the methodology it has learned. We close with a summary and concluding remarks in section~\ref{sec:conclusion}.

\section{Interpreting linear ML: Chern insulators}\label{sec:CI}

\begin{figure}
\includegraphics[scale=0.8]{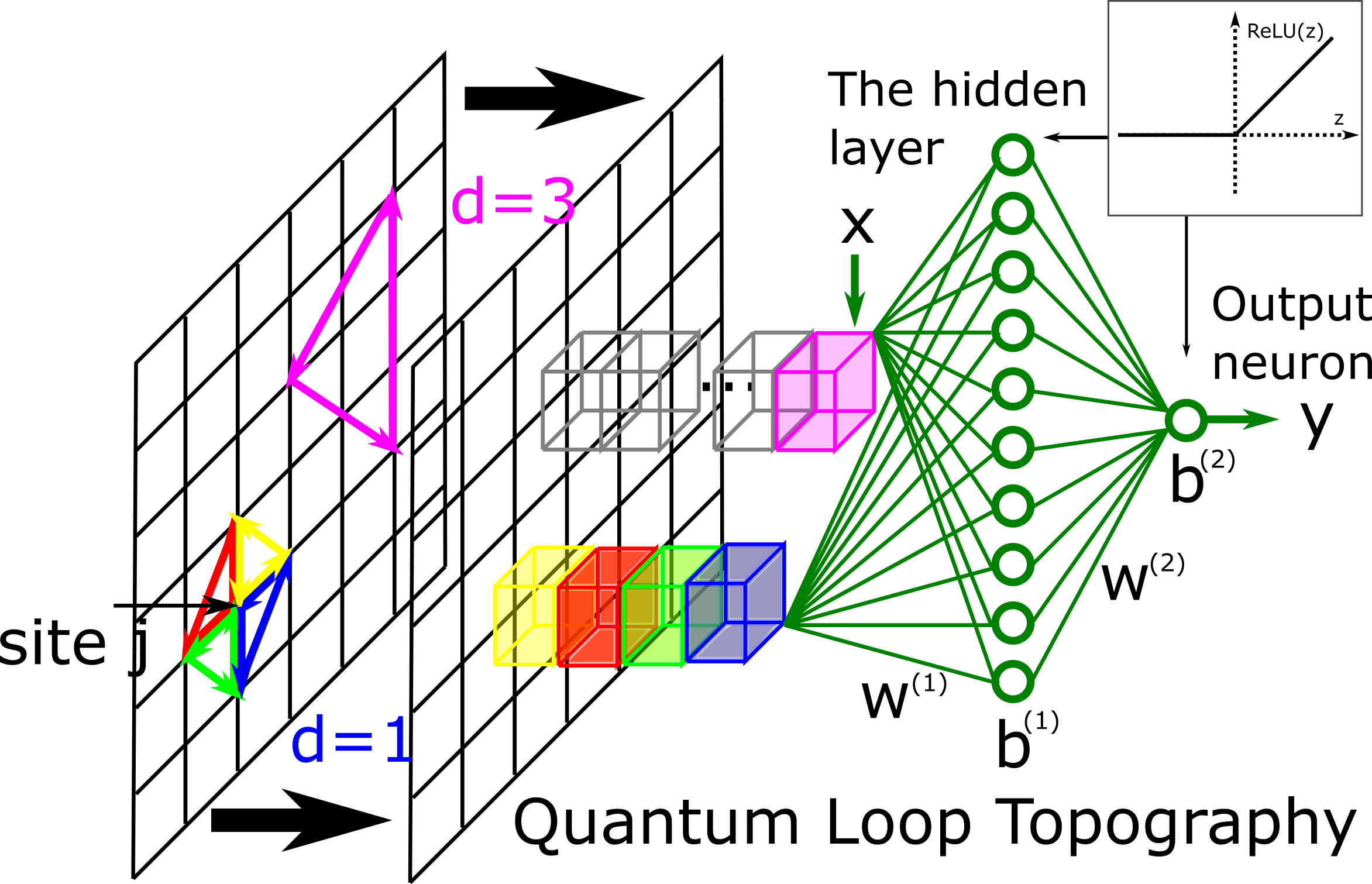}
\caption{Our supervised machine learning architecture for topological quantum Hall insulator consists of a quantum loop topography feature-selection layer and an ANN with a single hidden layer. The neurons are rectified linear units with $\mbox{ReLU}(z)=\max\left(z,0\right)$ as the nonlinear activation function, as illustrated in the inset. }
\label{fig:fig1}
\end{figure}

QLT for CI assigns a $D(d_c)$-dimensional vector of complex numbers to each lattice site $j$, thereby forming a quasi two-dimensional ``image''. The elements of the vector associated to the site $j$ are chained products such as
\begin{equation}
\tilde{P}_{jk}|_{\alpha}\tilde{P}_{kl}|_{\beta}\tilde{P}_{lj}|_{\gamma} \quad \left(\mbox{QLT for CI}\right),
\label{eq:tql}
\end{equation}
where $k$ and $l$ are two sites that form a triangle with  site $j$. In Eq.~\eqref{eq:tql}, each $\tilde{P}_{jk}|_{\alpha}\equiv \left\langle c^\dagger_j c_k \right\rangle_\alpha$ is a variational Monte Carlo sample of the two-point correlations associated with sites $j$, $k$ evaluated at Monte Carlo step $\alpha$, and $\beta$, $\gamma$ label different Monte Carlo steps. The length $D(d_c)$ of the vector is set by the total number of triangles anchored at the site $j$ with lateral distance $d\leq d_c$, where $d_c$ is the cutoff scale that can remain close to the lattice constant for a gapped system (see Fig.~\ref{fig:fig1}).

The above choice of QLT for CI is motivated by the characteristic response function that defines CI, the Hall conductivity for free fermion systems \cite{qlt2016}:
\begin{equation}
\sigma_{xy} = \frac{e^2}{h}\cdot\frac{1}{N} \sum_{\triangle jkl} 4\pi iP_{jk}P_{kl}P_{lj}S_{\triangle jkl},
\label{eq:sigmaxy}
\end{equation}
where $P_{ij}\equiv\langle c_i^\dagger c_j\rangle$ is the equal-time two-point correlation between sites $i$ and $j$, $S_{\triangle jkl}$ is the signed area of the triangle $jkl$, and $N$ is
the total number of sites \cite{Kitaev2006,Bianco2011}. Hence QLT in Eq.~\eqref{eq:tql} provides input that could contribute to the Hall conductivity, albeit with noisy single Monte Carlo instance data $\tilde{P}_{jk}|_{\alpha}\equiv \left\langle c^\dagger_j c_k \right\rangle_\alpha$. More importantly, since it is constructed from loops, QLT provides only gauge invariant data to the ANN.

We now we train an ANN using QLT from a model that exhibits a topological quantum phase transition (TQPT) between trivial insulator and Chern insulators \cite{Haldane1988},
and probe the ANN to interpret how it has learned. The model Hamiltonian is a tight-binding model on a two-dimensional square lattice \cite{qlt2016}:
\begin{eqnarray}
H(\kappa)&=&\underset{\vec{r}}{\sum}(-1)^{y} c_{\vec{r}+\hat x}^{\dagger}c_{\vec{r}} + [1+(-1)^{y} (1-\kappa)] c_{\vec{r}+\hat y}^{\dagger}c_{\vec{r}}\nonumber \\
&+& (-1)^{y}  \frac{i\kappa}{2}\left[c_{\vec{r}+\hat x + \hat y}^{\dagger}c_{\vec{r}}+c_{\vec{r}+\hat x - \hat y}^{\dagger}c_{\vec{r}}\right] + \mbox{h.c.}\ ,
\label{eq:sqham}
\end{eqnarray}
where $\vec{r} = (x,y)$ and $0\leq\kappa\leq1$ is a tuning parameter. The $\kappa=1$ limit is the $\pi$-flux square lattice model for a quantum Hall insulator with Chern number $C=1$, while the $\kappa=0$ limit reduces to decoupled two-leg ladders. $H(\kappa)$ interpolates between the quantum Hall insulator and the normal insulator with a TQPT at $\kappa=0.5$. We study a system of size $12\times 12$ lattice spacings.

Fig.~\ref{fig:fig1} shows the architecture of our single hidden layer ANN. The trained ANN with weights $w^{(1)}$, $w^{(2)}$ and biases $b^{(1)}$, $b^{(2)}$ transforms a QLT input vector $x$ into output $y$ given by
\begin{equation}
y= \sigma\left[\sum_{j}w_{j}^{(2)}\sigma\left(\sum_i w_{ji}^{(1)}x_i +b^{(1)}_j \right)+b^{(2)}\right]\ .
\end{equation}
$\sigma(z)$ is the non-linear activation function applied to the neurons, here taken as rectified linear units (RELU), i.e., with
\begin{equation}
\sigma(z)\equiv\mbox{ReLU}(z)=\max\left(z,0\right)\ .
\label{eq:ReLU}
\end{equation}
(The RELU choice has proven more efficient to train in the context of deep networks, and as well affords a simpler interpretation than the sigmoid function $S(z)=\left[1+\exp\left(z\right)\right]^{-1}$. used in Ref.\cite{qlt2016}.)

\begin{figure}
\includegraphics[scale=0.35]{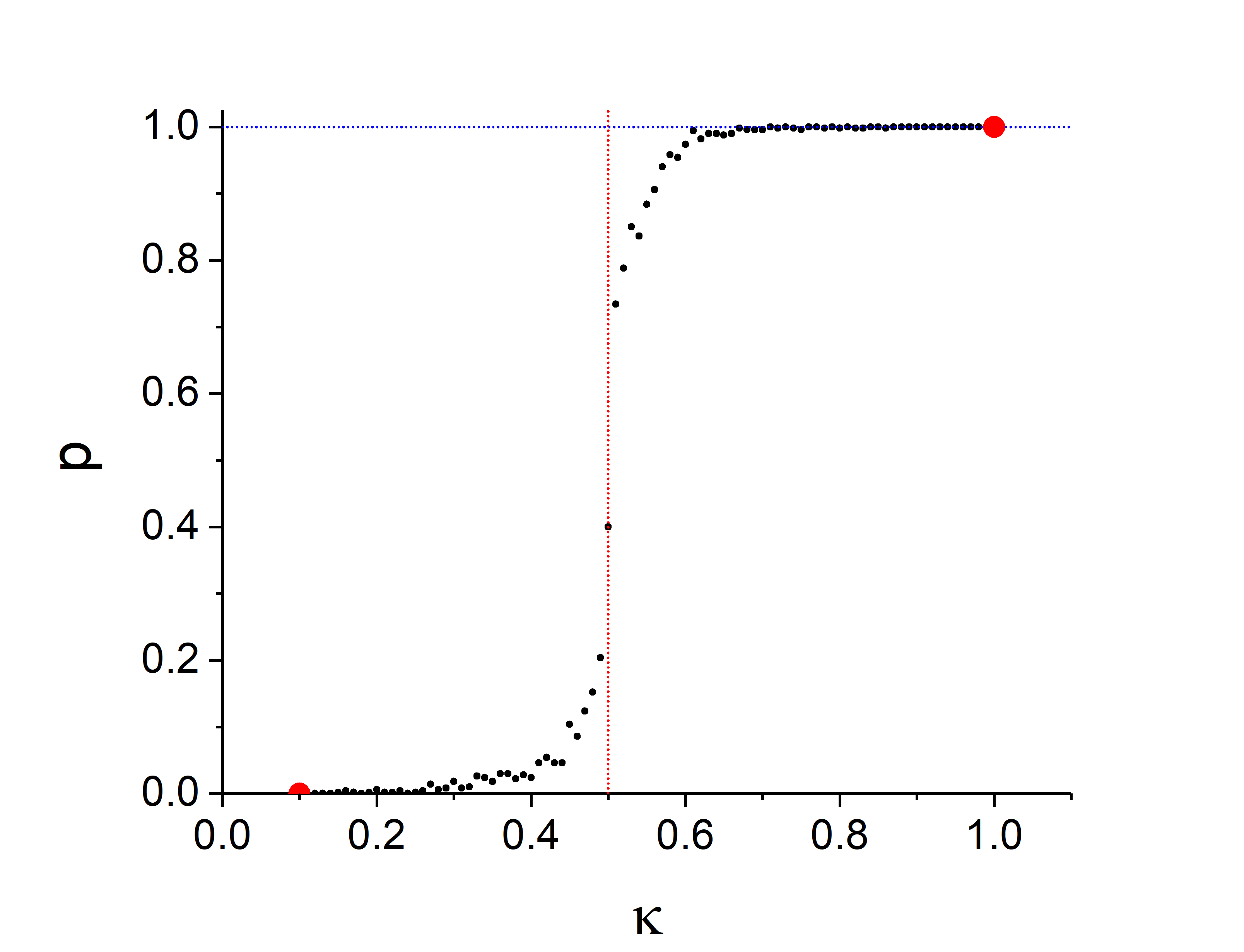}
\caption{After successful training on samples from both the quantum Hall insulator at $\kappa=1.0$ and trivial insulator phases at $\kappa=0.1$ (the filled circles in red at $(0.1, 0)$ and $(1, 1)$), the RELU ANN reliably provides consistent determination of the phase. $p$ is the probability that the ANN assigns a sample state to be in the quantum Hall phase. The vertical red line is the critical value $\kappa=0.5$ between the two phases.}\label{fig:fig2}
\end{figure}

Fig.~\ref{fig:fig2} shows the typical phase recognition by a successfully trained ANN. Here, the network confidence $p$ of the ground state being a Chern insulator at the given model parameter $\kappa$ is assessed by taking the average of the neural network outputs for 500 independent input samples at the given parameter value. An ANN trained at the two marked training points reliably detects the transition from trivial insulator to CI at $\kappa=0.5$.

\begin{figure*}[t]
  \centering
\subfigure[]{
\includegraphics[scale=0.3]{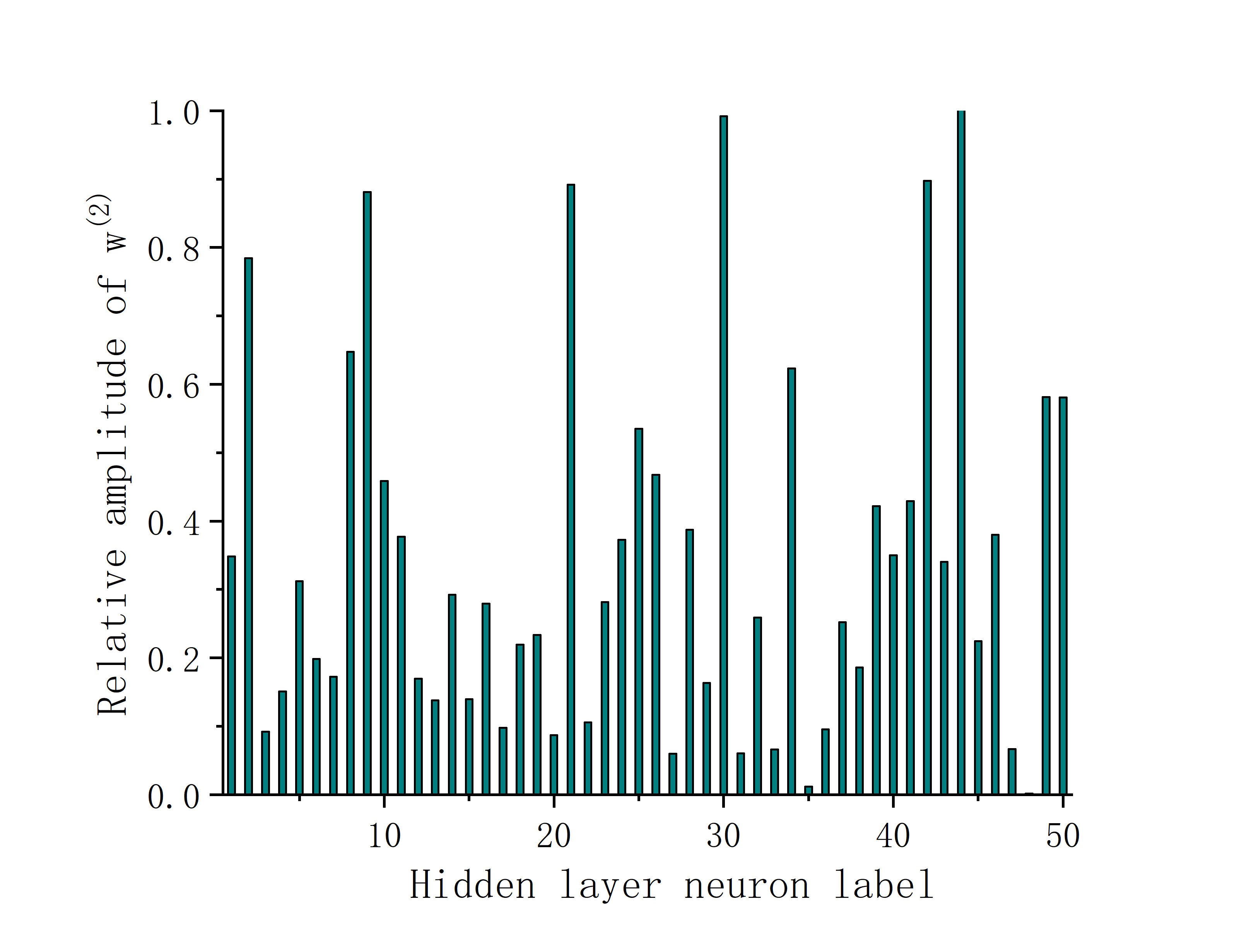}}
\subfigure[]{\includegraphics[scale=0.3]{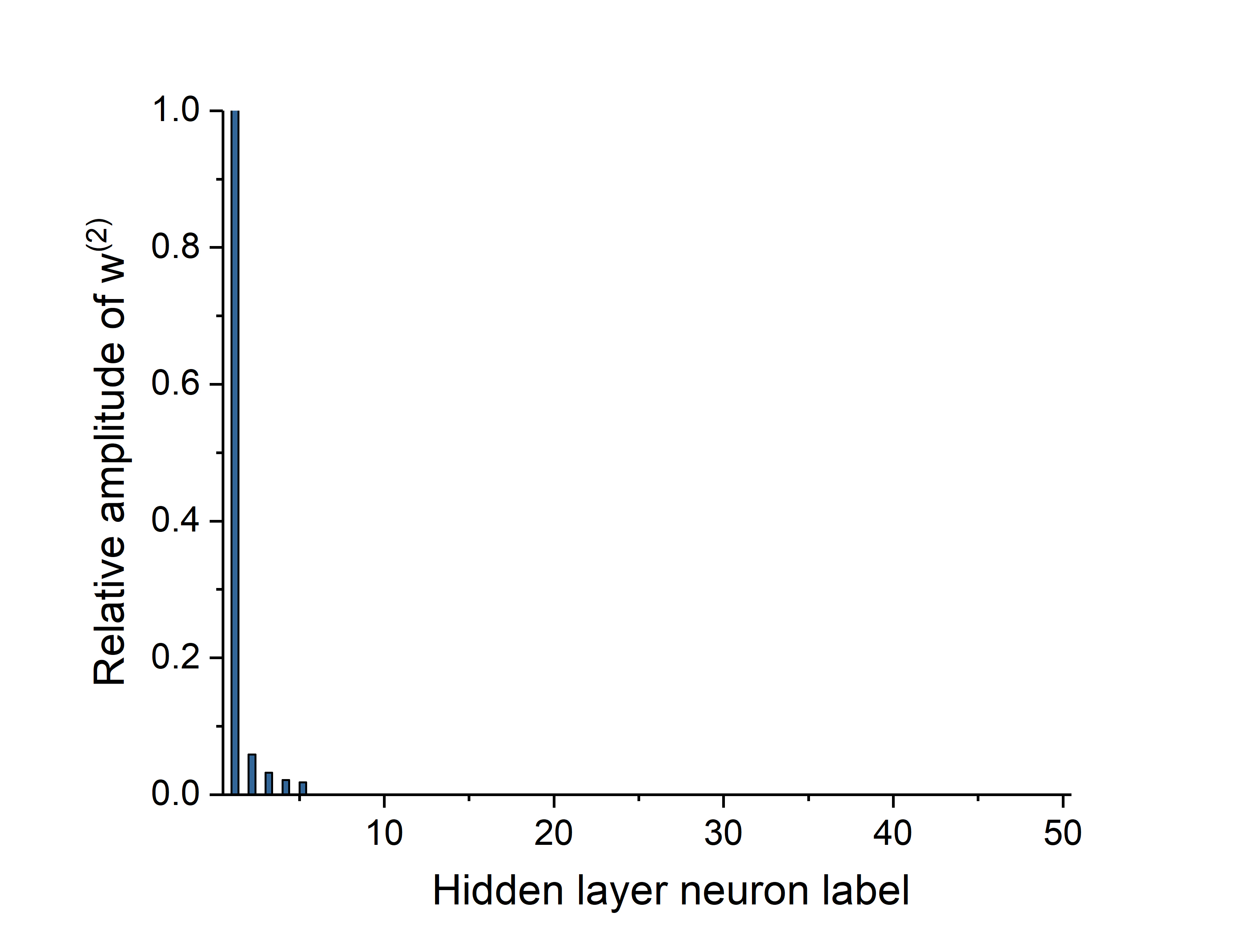}
}\\
\subfigure[]{
\includegraphics[scale=0.3]{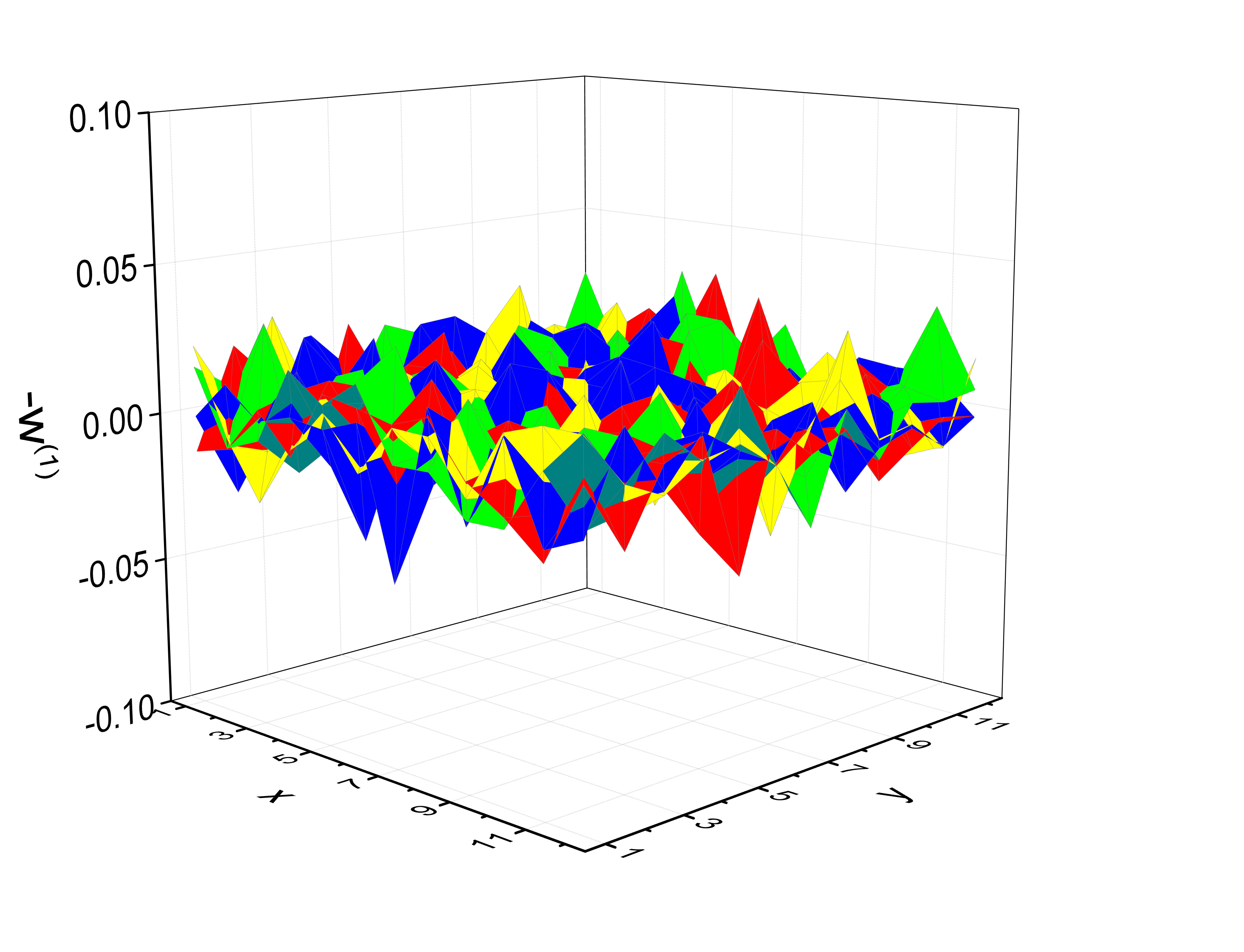}}
\subfigure[]{\includegraphics[scale=0.3]{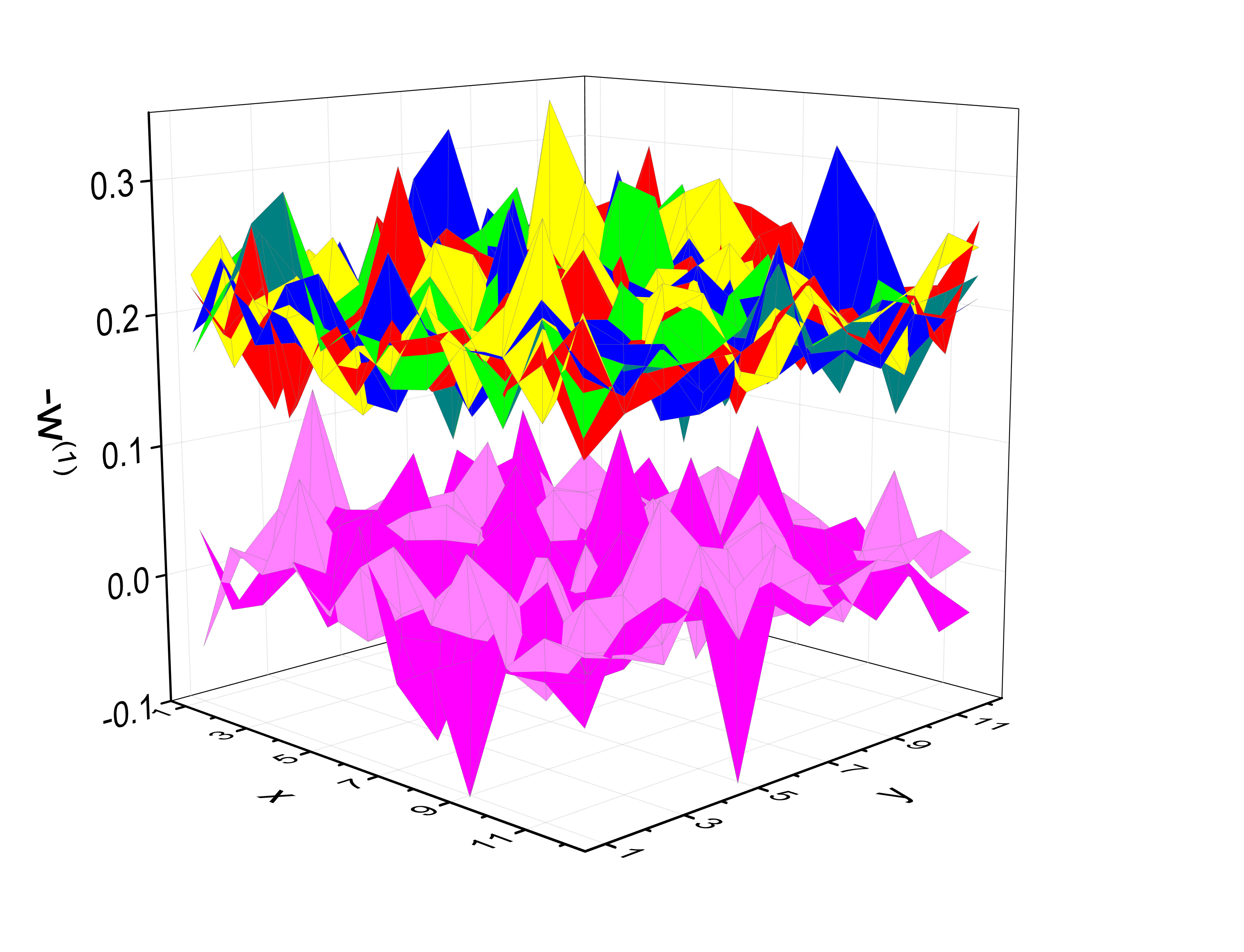}
}
\caption{(a,b) The distribution of the (absolute values of) the weights $w^{(2)}_j$ between the hidden layer neurons and the output neuron in an ANN with RELU neurons, (a) at random initialization before, and (b) after training. In the case shown, $j_{\rm max}$ is the very first neuron with $j=1$. (c,d) The distribution of weights $w^{(1)}_{j_{\rm max},i}$ associated with the input $i$ for the dominant neuron $j_{\rm max}$, (c) at random initialization before, and (d) after training. Here $i$ labels the type of triangle, the real and imaginary parts of the contributions, and the lattice position $\vec r = (x, y)$. The major contribution to the output comes from the \emph{imaginary parts} of the correlations from the four $d=1$ triangles (see Fig.~\ref{fig:fig1}), whose weights are depicted in red, yellow, green and blue, respectively. In comparison, the weights associated with other inputs are much closer to zero even after training, as illustrated by the $w^{(1)}_{j_{\rm max},i}$ distributions in magenta.}
\label{fig:fig3}
\end{figure*}

Our interpretation of a trained ANN begins with inspection of the final weights $w^{(2)}_j$ for each neuron $j$. Despite initialization with random fluctuations (Fig.~\ref{fig:fig3}(a)), we often find the $w^{(2)}_j$ to be largely concentrated on a single hidden layer neuron, which we label $j_{\rm max}$, see Fig.~\ref{fig:fig3}(b). With the ANN output determined by this single neuron, we can ignore the rest of the hidden layer neurons, and trace the firing of the output neuron back to the firing condition of $j_{\rm max}$, in turn encoded in the weights $w^{(1)}_{j_{\rm max}, i}$ and bias $b^{(1)}_{j_{\rm max}}$. Here $i=\left(\vec r, t\right)$ labels the inputs according to their lattice sites $\vec r$ and the triangles $t \in[1,D(d_c)]$. The real and imaginary parts are also treated as separate inputs. Fig.~\ref{fig:fig3}(c) and (d) show the distribution of selected $w^{(1)}_{j_{\rm max}, i}$ for each lattice site $\vec{r}$ at random initialization and after the training. By inspection of $w^{(1)}_{j_{\rm max},i}$ for all $i$'s we find that the neuron $j_{\rm max}$ ended up weighting as most significant the \emph{imaginary parts} of $\tilde{P}_{jk}|_{\alpha}\tilde{P}_{kl}|_{\beta}\tilde{P}_{lj}|_{\gamma}$ coming from the smallest triangles $jkl$, namely, $j={\vec r}$, $k={\vec r\pm \hat x}$, and $j={\vec r \pm \hat y}$. Moreover, $w^{(1)}_{j_{\rm max}, i}$ is approximately evenly distributed across all four $d=1$ triangles and across all real-space positions $\vec r$, as shown by the four colors for the four triangles in Fig.~\ref{fig:fig3}(d). For all of the other inputs, including all real parts, the associated weights $w^{(1)}_{j_{\rm max}, i}$ are close to zero, as shown in magenta in Fig.~\ref{fig:fig3}(d).

The observation of significant separation between the weights $w^{(1)}_{j_{\rm max},i}$ of the smallest triangles and the rest implies that we can approximate the firing condition by concentrating on those smallest triangles. Moreover, since the weights for the smallest triangles cornered at each site are roughly the same, we can approximate the firing condition for the Chern Hall insulator by the following criterion:
\begin{equation}
w^{(2)}_{j_{\rm max}} \max[\ \bar w^{(1)}_{j_{\rm max}}\!\!\!\sum_{i={\vec r,\pm \hat x, \pm \hat y}} \mbox{Im} P_{\vec r\pm \hat y,\vec r}P_{\vec r,\vec r\pm \hat x}P_{\vec r\pm \hat x, \vec r\pm \hat y}+b^{(1)}_{j_{\rm max}},0]+b^{(2)}>0
\label{eq:ciresult}
\end{equation}
In the above, we have replaced the $w^{(1)}_{j_{\rm max},i}$ over all positions $\vec r$ and four triangles (labeled by the relative position of the other two vertices $\pm \hat x$ and $\pm \hat y$) with their average $\bar w^{(1)}_{j_{\rm max}}$.
Reading off the weights and biases from a learned ANN and inserting their values, $\bar w^{(1)}_{j_{\rm max}}=-0.208$, $w^{(2)}_{j_{\rm max}}\sim -4.84$, $b^{(1)}_{j_{\rm max}}\sim 3.73$, and $b^{(2)}\sim 9.03$ into Eq.~\eqref{eq:ciresult} yields
\begin{eqnarray}
-4.84\times \max[-0.208\sum_{d_{\triangle{jkl}}=1} \mbox{Im} P_{jk}P_{kl}P_{lj}+3.73,0]+9.03&>&0\nonumber \\
\Leftrightarrow \frac{4\pi}{N} \sum_{d_{\triangle jkl}=1} -\mbox{Im}P_{jk}P_{kl}P_{lj}/2 > 0.4,
\end{eqnarray}
where $N=L^2=144$. Considering $S_{\triangle jkl}=1/2$ for the $d_{\triangle jkl}=1$ triangles, the above criterion implies that the ANN relied on the imaginary part of the QLT input associated with the smallest triangles, and diagnosed the system to be a Chern insulator when their contribution to the Chern number was substantial. This is a reasonable and efficient diagnosis, given the exact formula Eq.~(\ref{eq:sigmaxy}) for the invariant in the position basis.

Our successful interpretation of the ANN's learning in the case of the Chern insulator was thus enabled by two aspects of our approach: (i) the QLT was effective in providing the relevant features, and (ii) the exact topological invariant was known in the local basis and so guided our interpretation. The effectiveness of the QLT is reflected in the ANN learning a linear function based on a single $j_{\rm max}$ neuron.

\section{Non-linear ML: the two-dimensional $\mathbb{Z}_2$ topological insulators}\label{sec:QSH}

\begin{figure}
\subfigure[]{
\includegraphics[scale=2.0]{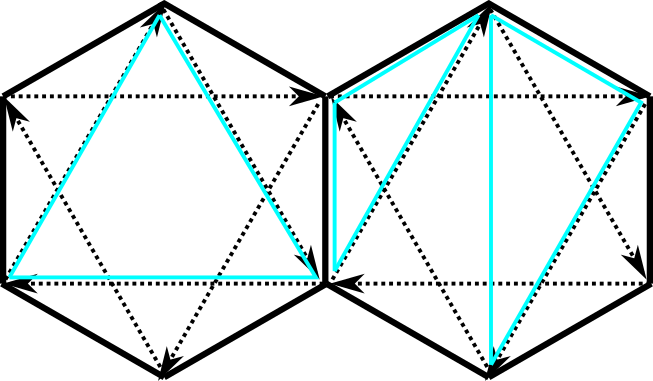} }\\
\subfigure[]{
\includegraphics[scale=0.3]{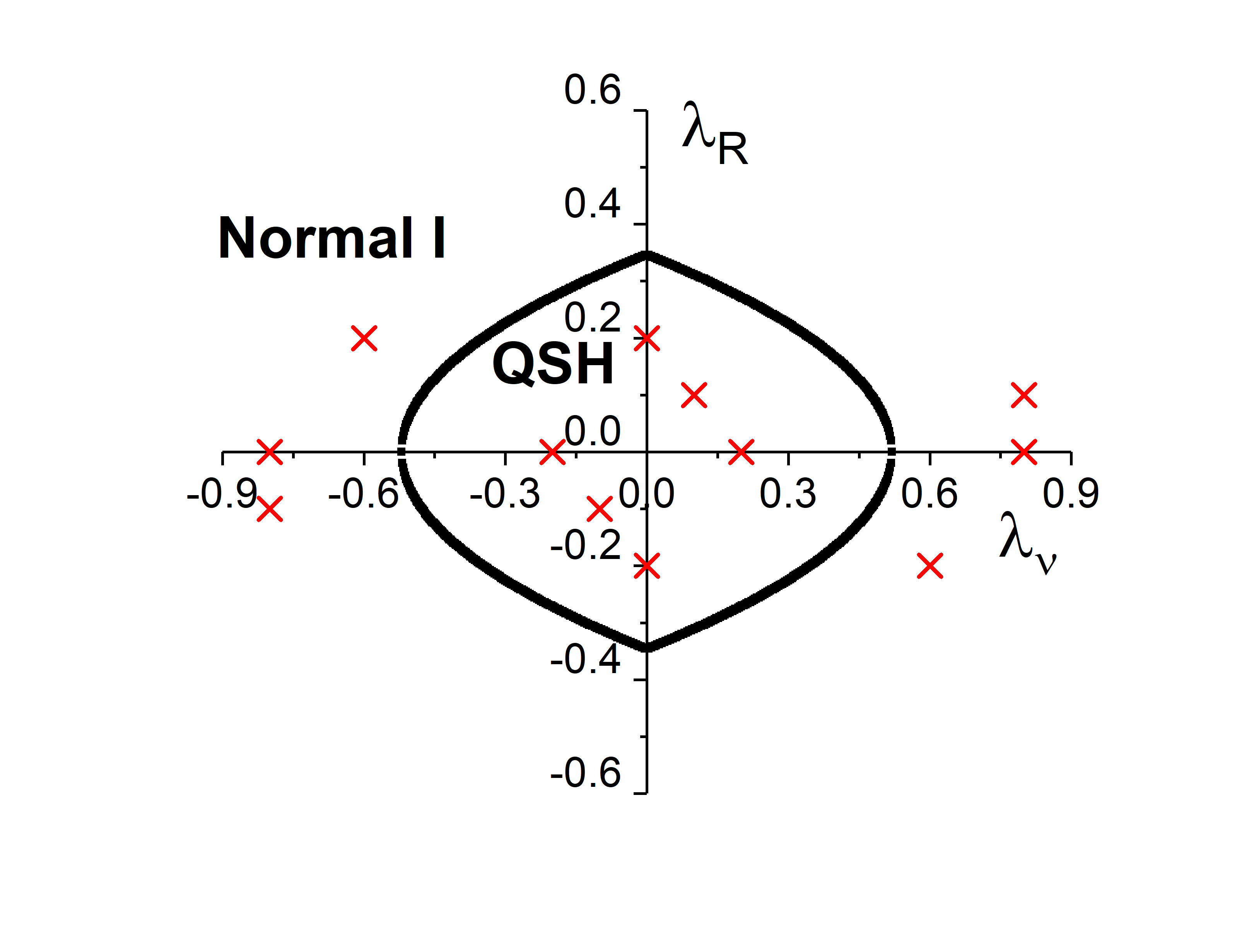}}
\caption{Upper panel: the honeycomb lattice for the model Hamiltonian in Eq.~\ref{eq:qsheham}. The sign $\nu_{ij}$ of the spin orbit interaction between the next nearest neighbors are positive along the arrows and negative against the arrows. The three smallest types of triangles are illustrated in cyan. Lower panel: the phase diagram of the tight-binding model in Eq.~\ref{eq:qsheham} with $t=1$ and $\lambda_{SO}=0.1$. The red crosses indicate the parameters used for the supervised machine learning training set.}
\label{fig:qshelattice}
\end{figure}

We now turn to a topological insulator with no known expression for the topological invariant in the position basis: the time-reversal-invariant $\mathbb{Z}_2$ TI in two dimensions, also known as the quantum spin Hall insulator, with a quantized spin Hall conductance and a vanishing charge Hall conductance. The characteristic $\mathbb{Z}_2$ topological invariant is only known as a loop-integral
\begin{equation}
I=\frac{1}{2\pi i}\oint d\vec{k} \cdot \partial_{\vec k} \log[\mbox{Pf}\left(\vec k\right)+i\delta],
\label{eq:Z2}
\end{equation}
of the phase winding of the Pfaffian $\mbox{Pf}\left(\vec k\right) = \mbox{Pf} \left[\left\langle u_n(\vec k) | \Theta | u_m(\vec k) \right\rangle\right]$ over a contour in momentum space enclosing half the Brillouin zone \cite{QSHE2005}, where $n$ and $m$ are band labels, and $\Theta$ is the time reversal operator. A position-basis expression for the $\mathbb{Z}_2$ index $I$, the counterpart of Eq.~\ref{eq:sigmaxy} for the Chern number, is not known in general.

In the presence of spin $s_z$ conservation, a $\mathbb{Z}_2$ TI is equivalent to two copies of Chern insulators, with $\sigma^{\uparrow}_{xy}=1$ for the spin up ($s_z=\uparrow$) electrons and an anti-chiral quantum Hall insulator  $\sigma^{\downarrow}_{xy}=-1$ for the spin down electrons($s_z=\downarrow$). Hence Eq.~\eqref{eq:sigmaxy} for each spin component will serve as a position-basis expression for the $\mathbb{Z}_2$ index. It is known, however, that the $\mathbb{Z}_2$ TI state is well-defined through the momentum space expression for the $\mathbb{Z}_2$ index Eq.~\eqref{eq:Z2}, even when the Rashba spin-orbit coupling breaks $s_z$ conservation.

We turn to the physical response of an effective spin-current to a transverse electric field: the spin-Hall conductivity. \textcite{Niu2005} introduced the effective spin current in spin-orbit coupled systems, in the absence of average torque, as a time-derivative of the spin-displacement operator: $\hat{\mathcal{J}}_s=\frac{d\hat{r}\hat{s}_z}{dt}=1/\hbar[H,\hat{r}\hat{s}_z]$. The flat-band Hamiltonian is defined as $\hat{H}'=1-\hat{P}$, where  $\hat{P}$ is the projection operator onto the valence band of $H$. $\hat{H}'$ is adiabatically connected to the model Hamiltonian $H$, and thus shares the topological properties such as the spin-Hall conductivity and the $\mathbb{Z}_2$ topological index. According to the Kubo formula, the spin-Hall conductivity of $H'$ is
\begin{eqnarray}
& &\mbox{tr} P \left[P, x \hat{s}_z\right] \left[P, y\right] \\\nonumber
&=&\sum_{\triangle jkl, s^z_j, s^z_k, s^z_l} P_{j,s^z_j;k,s^z_k}P_{k,s^z_k;l,s^z_l}P_{l,s^z_l;j,s^z_j}\left( \hat{s}^{z}_{j} x_j - \hat{s}^{z}_{k} x_k  \right)\left(y_k - y_l\right)
\end{eqnarray}
where $P_{j,s^z_j;k,s^z_k}\equiv \langle c^\dagger_{j,s^z_j;} c_{k,s^z_k}\rangle $ are the two-point correlators of $H$ and the summation is over all triangles, with vertices $j$, $k$ and $l$. Since there is no spin-quantization direction in the presence of Rashba spin-orbit coupling, we propose to use the following QLT to probe the quantum spin Hall (QSH) effect:
\begin{eqnarray}
& &s^{x}_{j}\tilde{P}_{jk}|_{\alpha}\tilde{P}_{kl}|_{\beta}\tilde{P}_{lj}|_{\gamma} \nonumber\\
\left(\mbox{QLT for QSH}\right): \quad  & &s^{y}_{j}\tilde{P}_{jk}|_{\alpha}\tilde{P}_{kl}|_{\beta}\tilde{P}_{lj}|_{\gamma} \nonumber\\
& & s^{z}_{j}\tilde{P}_{jk}|_{\alpha}\tilde{P}_{kl}|_{\beta}\tilde{P}_{lj}|_{\gamma}.
\label{eq:QLT-SQH}
\end{eqnarray}
As in the last section, $\tilde P$'s are to be evaluated at independent Monte Carlo steps, and we focus only on the smallest triangles $jkl$, which should account for the major contributions, due to the exponentially decaying correlations in a gapped system.

We now employ the QLT in Eq.~\eqref{eq:QLT-SQH} to train an ANN to recognize the $\mathbb{Z}_2$ TI phase in the two-parameter phase space of the Kane-Mele model of Ref~\cite{QSHE2005}:
\begin{eqnarray}
H&=&t\sum_{\left\langle ij\right\rangle} c^\dagger_i c_j + i \lambda_{SO} \sum_{\left\langle\left\langle ij\right\rangle \right\rangle} \nu_{ij} c^\dagger_i S^z c_j + i \lambda_{R} \sum_{\left\langle ij \right\rangle}  c^\dagger_i \left(\vec{s}\times\hat d_{ij}\right)_z c_j \nonumber \\
& &+ \lambda_\nu \sum_{\left\langle i\right\rangle} \xi_i c^\dagger_i c_i\ ,
\label{eq:qsheham}
\end{eqnarray}
where $t=1$ is the nearest neighbor hopping amplitude, $\lambda_{SO}=0.1$ is the spin-orbit coupling between the next nearest neighbors, the sign $\nu_{ij}=\pm 1$ depends on whether the direction is along or against the arrow (see Fig.~\ref{fig:qshelattice} upper panel), $\lambda_\nu$ is a staggered potential, and $\lambda_R$ is the Rashba term. For $\lambda_R=0$, $s^z$ is a good quantum number, and the model reduces to two independent copies of quantum Hall insulators \cite{Haldane1988}. In the presence of a finite $\lambda_R$ Rashba term, however, $s^z$ is no longer a good quantum number, and there is no longer a conserved spin. For small ratios of $\lambda_R/\lambda_{SO}$ and $\lambda_\nu/\lambda_{SO}$, the model Eq.~\eqref{eq:qsheham} is known to realize a two-dimensional $\mathbb{Z}_2$ TI, and otherwise a normal insulator \cite{QSHE2005}. The exact phase diagram in Fig.~\ref{fig:qshelattice} lower panel is obtained through an explicit evaluation of the $\mathbb{Z}_2$ index given in Eq.~\eqref{eq:Z2}. Here we attempt to reproduce the phase diagram using QLT-based machine learning.

\begin{figure}
\includegraphics[scale=0.3]{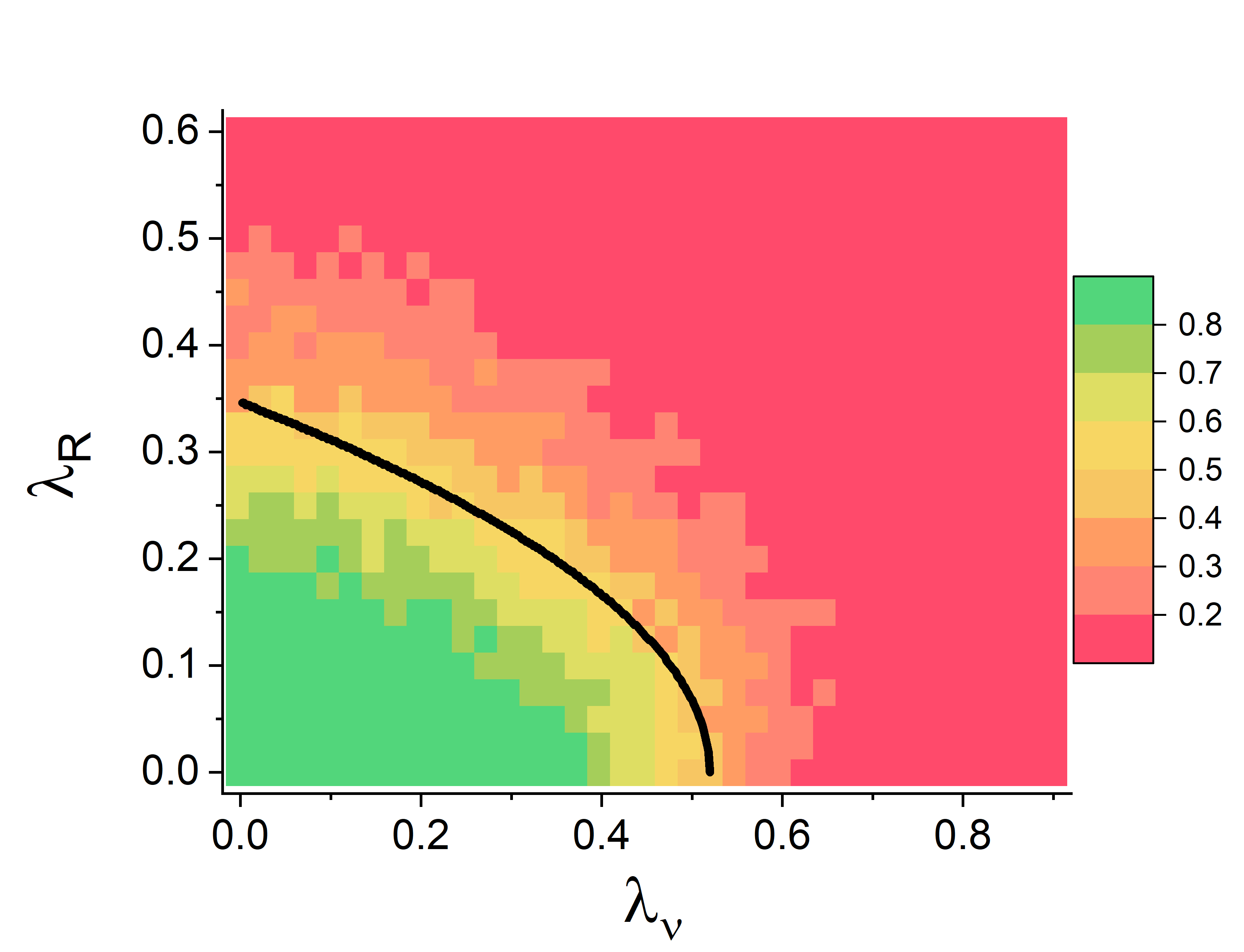}
\caption{After supervised machine learning, the ANN predicted likelihood of the Hamiltonian in Eq.~\ref{eq:qsheham} being a $\mathbb{Z}_2$ TI in the $[\lambda_\nu,\lambda_R]$ parameter space, the upper right quadrant of the phase diagram in Fig.~\ref{fig:qshelattice} lower panel. }
\label{fig:fig6}
\end{figure}

We consider variational Monte Carlo samples of $s^{x}_{j} P_{jk} P_{kl} P_{lj}$, $s^{y}_{j} P_{jk} P_{kl} P_{lj}$ and $s^{z}_{j} P_{jk} P_{kl} P_{lj}$ over the three smallest types of triangles (see Fig.~\ref{fig:qshelattice} upper panel) and feed the corresponding QLT inputs into an ANN with a single hidden layer of RELU, as in the previous section. The training set consists of QLTs obtained from exact wave functions at the training points marked in Fig.~\ref{fig:qshelattice} lower panel. To avoid approximate conservation of spin $s^z$, we randomly cycle $s^x$, $s^y$, and $s^z$ of the training samples during supervised machine learning. The trained ANN is then applied to QLT samples obtained from the phase space between $\lambda_\nu \in [0, 0.9]$ and $\lambda_R \in [0, 0.6]$ to assess the likelihood that the input belongs to a $\mathbb{Z}_2$ TI.

The resulting phase diagram obtained by the ANN is shown in Fig.~\ref{fig:fig6}, and is consistent with the exact phase diagram. The benefit of using the QLT-based local input is that this approach allows the investigation of systems with disorder. With the original definition of the $\mathbb{Z}_2$ invariant requiring a momentum space integral, there was no framework to assess whether a realistic system with disorder is a $\mathbb{Z}_2$ TI (although due its topological nature one would expect the $\mathbb{Z}_2$ TI to be immune to small disorder). Our success in using the QLT-based machine learning to recognize the $\mathbb{Z}_2$ TI paves the way for studying realistic models with disorder to discover more $\mathbb{Z}_2$ TI in nature.

\begin{figure}
\includegraphics[scale=0.3]{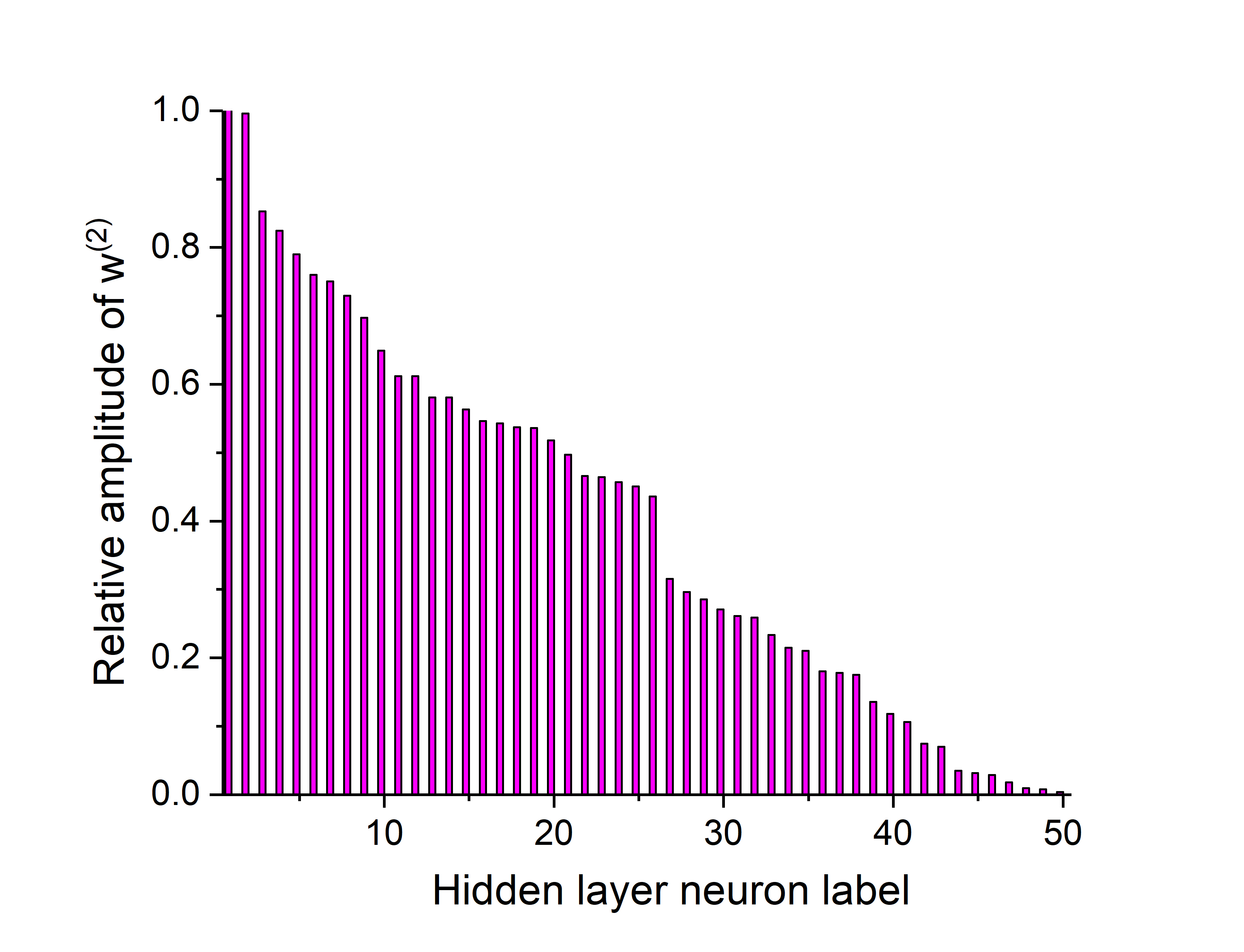}
\caption{The distribution of the (absolute values of) the weights $w^{(2)}_j$ between the hidden layer neurons and the output neuron in an ANN with RELUs. The decision boundary for the $\mathbb{Z}_2$ TI involves the interplay between multiple hidden layer neurons. }
\label{fig:w2forqsh}
\end{figure}

To interpret the ANN's learning from the QLT for the $\mathbb{Z}_2$ TI Eq.~\eqref{eq:QLT-SQH}, we again plot the modulus of $w^{(2)}_{j}$ between the hidden layer RELUs and the output neuron. We now find all the neurons to have substantial weight in the trained ANNs (see Fig.~\ref{fig:w2forqsh}). Contrasting the weight distribution in Fig.~\ref{fig:w2forqsh} for $\mathbb{Z}_2$ TI with that for the CI (see the right column of Fig.~\ref{fig:fig3}), we see that the ANN for the QSH learns a more complex function of the QLT input. The earlier concentration of $w^{(2)}_{j}$ to one neuron $j_{\rm max}$ for the CI means that the ANN effectively expressed a linear function of the QLT as threshold for that neuron, because the QLT features were already so effective for the CI. That the ANN forms a non-linear function with the QLT inputs for the QSH implies that the position-based expression for the $\mathbb{Z}_2$ invariant is not a simple linear combination of the QLT Eq.~\eqref{eq:QLT-SQH}. This further motivates the use of the ANN-based approach for the QSH. It is moreover plausible that we can gain insight into the presently unknown position-based expression for the $\mathbb{Z}_2$ invariant, by studying the weights and biases of the function learned by the ANN.\footnote{Please contact the authors for preliminary results.} It might also be possible to adapt conventional machine learning methods such as the Lasso \cite{lasso}, or more general $L^1$ regularizations, to enforce sparsity as in Fig.~\ref{fig:fig3}b, and thereby facilitate interpretability.

\section{Interpreting non-linear ML: the $\mathbb{Z}_2$ lattice gauge theory}\label{sec:QSL}

In this section, we turn to strongly interacting models of lattice gauge theory. Previously, ANN-based machine learning has been used to improve the efficiency of lattice QCD simulations \cite{Shanahan2018} and detected phase boundaries of topological quantum phase transitions \cite{Melko20161, MelkoIntp2017, FrankMLZ2}, using the mapping between the $T=0$ quantum problem in two dimensions and the lattice gauge theory in three dimensions.
Here we revisit the ANN-based phase diagram of Ref.~\cite{FrankMLZ2}, where two of us successfully trained an ANN to recognize the $\mathbb{Z}_2$ quantum spin liquid phase, the deconfined phase for the corresponding three-dimensional $\mathbb{Z}_2$ lattice gauge theory. The goal is to interpret what the ANN learns from the semi-local QLT training data. Specifically, the question is how the ANN, using the behavior of small loops as features, can detect a confining phase, ordinarily signaled by the area vs.\ length behavior of large loops, as explained below.\footnote{We thank Eduardo Fradkin for first raising this question to us.}

For simplicity, we consider the $\mathbb{Z}_2$ lattice gauge theory given by the Hamiltonian on a three dimensional cubic lattice:
\begin{equation}
\beta H_{3D}=-\lambda_b \sum_j S_j -\lambda_p\sum_p \prod_{j\in p} S_j
\end{equation}
where $S_j = \pm 1$ lives on the bonds of the cubic lattice, and $p$ denotes the square plaquettes. For small values of $\lambda_b$, the system has a phase transition at the critical value $\lambda_p\sim 0.76$ between a deconfined phase at large $\lambda_p$ and a confined phase at small $\lambda_p$.

\begin{figure}
\includegraphics[scale=0.40]{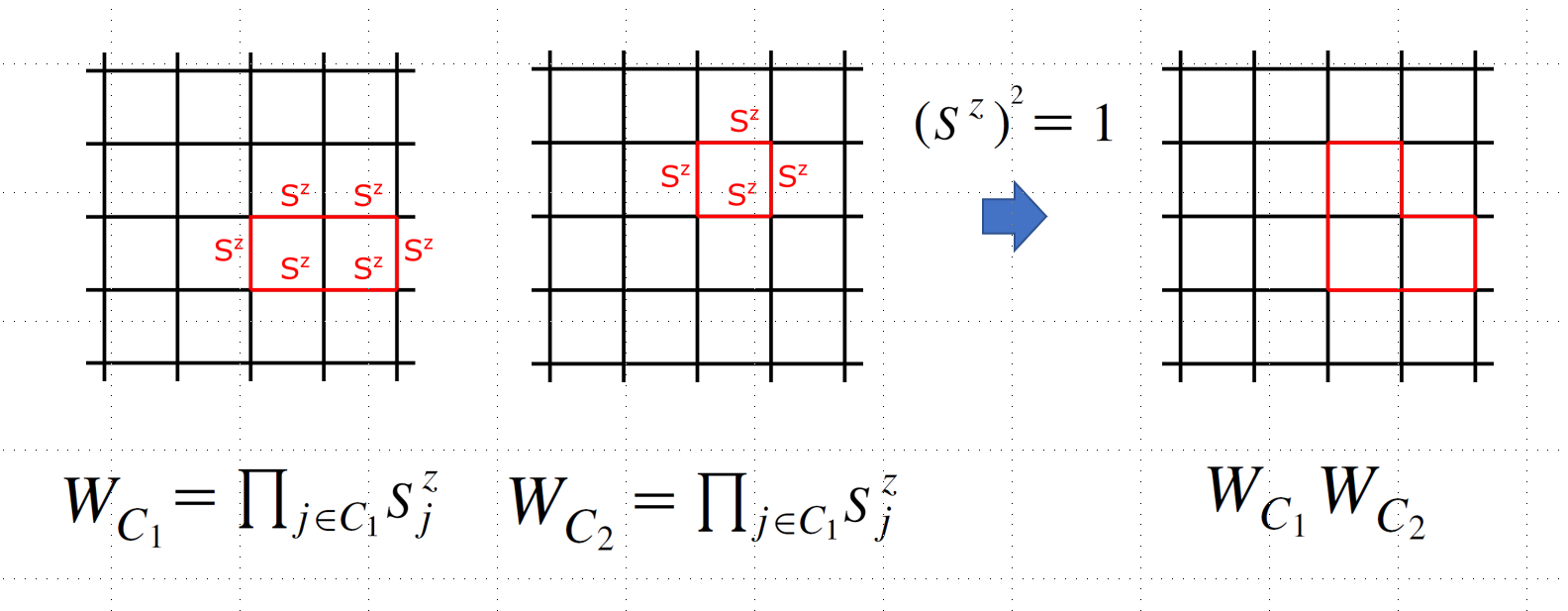}
\caption{Wilson loops in the $\mathbb{Z}_2$ lattice gauge theory correspond to product of spin $s^{z}$ operators around a closed string. The product of smaller loops can give rise to larger loops.}
\label{fig:looproduct}
\end{figure}

An important physical concept in the lattice gauge theory is the Wilson loop: the path-ordered gauge field integrated around a closed loop. Since the Wilson loops are gauge invariant, they provide meaningful measurements of the lattice gauge theory and a natural dataset for machine learning \cite{Shanahan2018}. For the $\mathbb{Z}_2$ lattice gauge theory, the Wilson loop around a given loop $C$ is defined as
\begin{equation}
W_C=\prod_{j\in C} S_j\ .
\end{equation}
The deconfined phase of the lattice gauge theory is not distinguished by any broken symmetries, but rather by the limiting behavior of the Wilson loop in the thermodynamic limit: in the confined phase, the expectation value of the Wilson loop decays exponentially $\left\langle W_C\right\rangle \propto \exp(-A_C)$ as a function of the area $A_C$ enclosed by the loop $C$, whereas in the deconfined phase, $\left\langle W_C\right\rangle \propto \exp(-l_C)$ decays only as the length $l_C$ of the loop $C$ \cite{Fradkin1979}.

\begin{figure}
\includegraphics[scale=1.0]{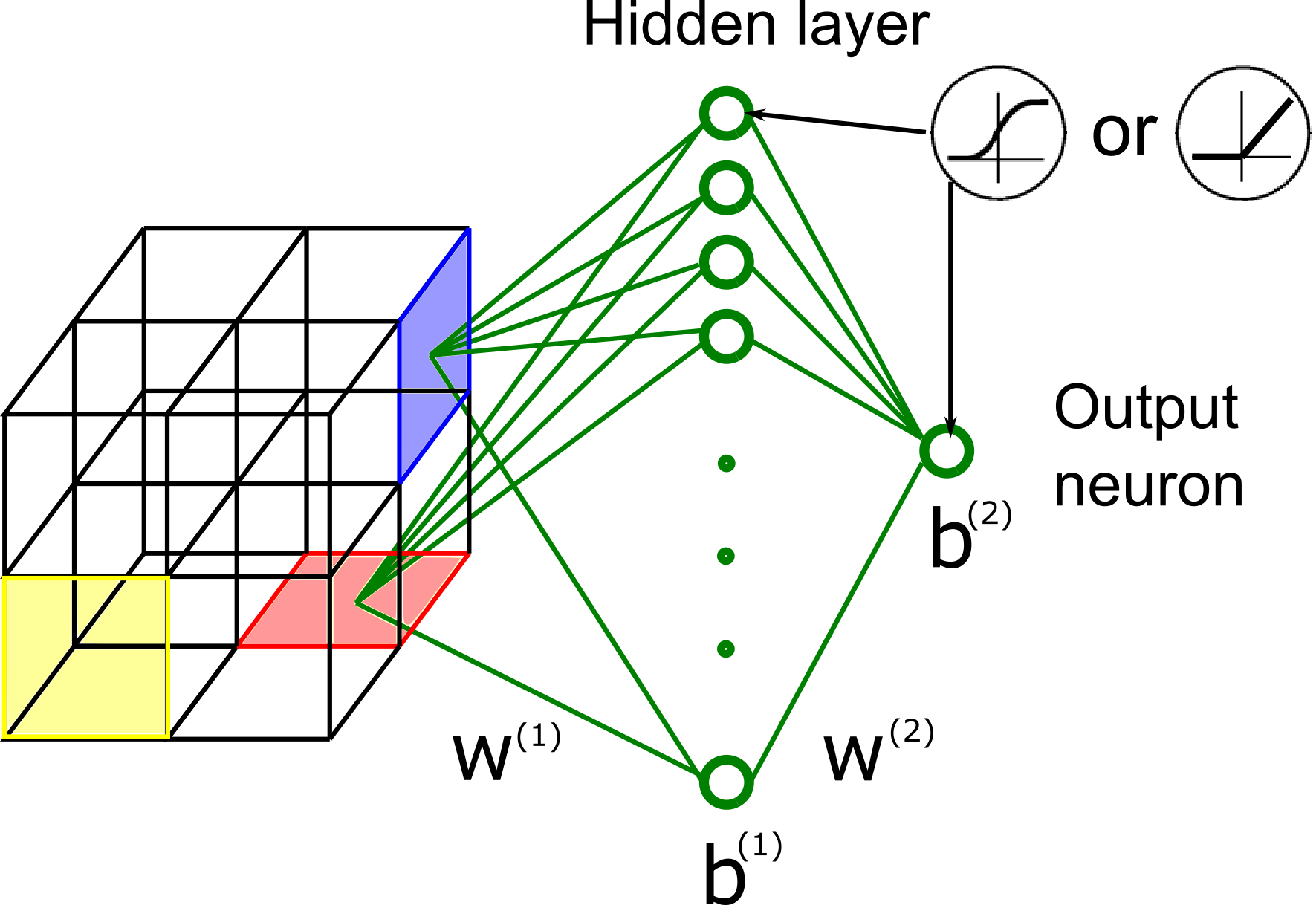}
\caption{The application of supervised machine learning on the $\mathbb{Z}_2$ lattice gauge theory uses Monte Carlo samples of the smallest Wilson loops, the plaquettes, as inputs to the ANN.
The ANN consists of neurons with RELU activation functions.}
\label{fig:fig5}
\end{figure}

In what follows, we will take advantage of the Abelian nature of $\mathbb{Z}_2$, which permits small Wilson loops to fuse in a unique channel and form larger ones, consisting of the products of the smaller ones, as in Fig.~\ref{fig:looproduct}. In a single Monte Carlo snapshot, the sampled values of all Wilson loops can thus be obtained from those of the smallest Wilson loops, $W_p=\prod_{j\in p} S_j$ around the square lattice plaquettes. In accord with this observation, we use the classical Monte Carlo samples of the smallest Wilson loops $W_p$ on the $L=12$ dual lattice as the inputs to the ANN, see Fig.~\ref{fig:fig5}.

\begin{figure}
\includegraphics[scale=0.3]{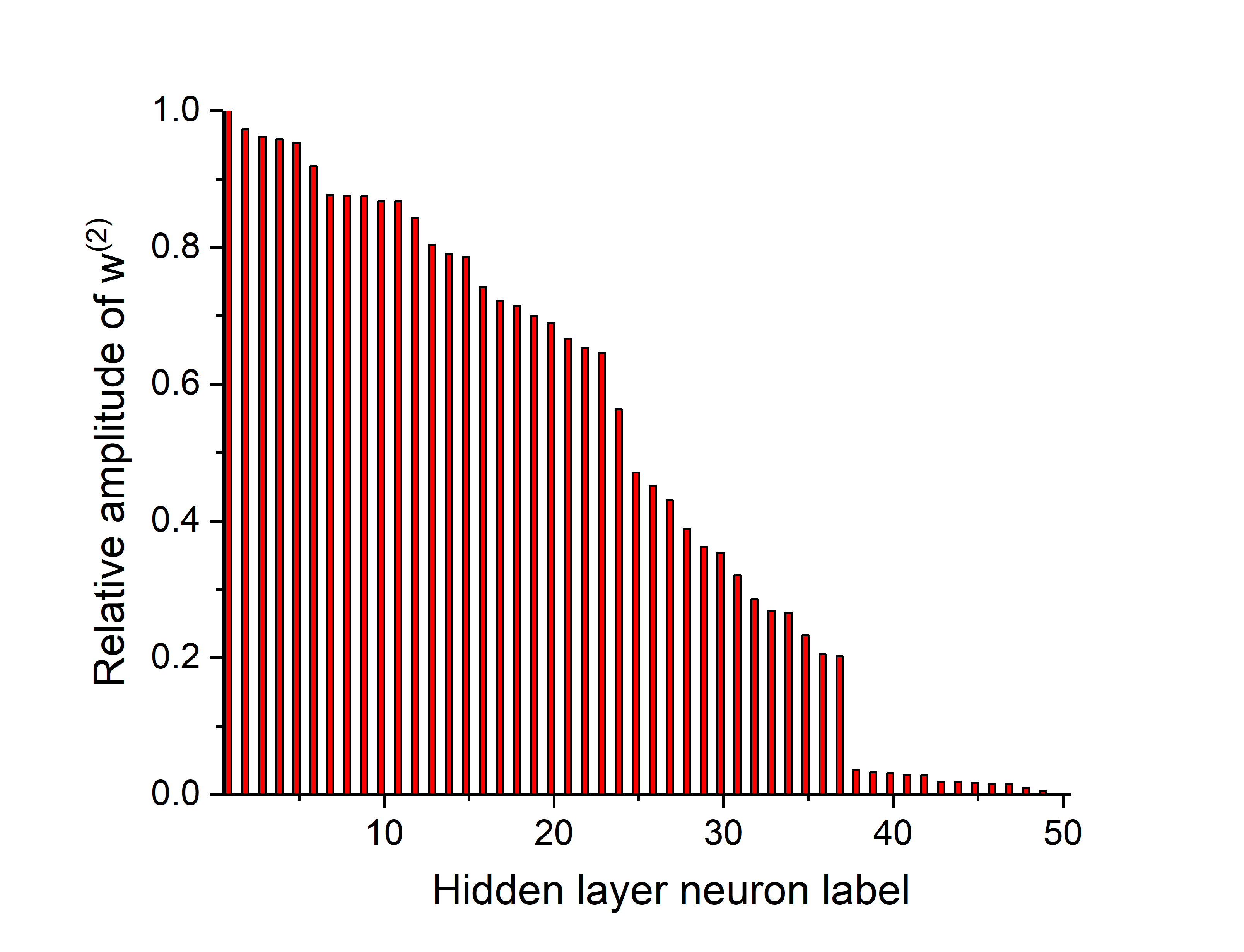}
\caption{The distribution of the absolute values of the weights between the hidden layer neurons and the output neuron. The decision boundary for the $\mathbb{Z}_2$ lattice gauge theory ($\mathbb{Z}_2$ topological order) involves the interplay between multiple hidden layer neurons.}
\label{fig:Z2-lattice-w2}
\end{figure}

We use normalized Monte Carlo samples of the set of $W_p$ for the two distinctive phases above ($\lambda_p =0.83$) and below ($\lambda_p =0.68$) the critical value as the training set, with $\lambda_b=0.1$, to perform the supervised learning. In ref.~\cite{FrankMLZ2}, the phase diagram mapped out for the $\lambda_b/\lambda_p$ plane by the optimized ANN was found to be in good agreement with the value of $T_c$ determined by finite-size scaling of Monte Carlo data, giving confidence in that phase diagram. As in the earlier sections, we use a shallow ANN with RELU neurons (see Fig.~\ref{fig:fig5}). An inspection of an ANN trained to correctly distinguish phases of the $\mathbb{Z}_2$ lattice gauge theory ($\mathbb{Z}_2$ topological order) shows multiple hidden layer neurons to participate. This is shown in the distribution of weights $w_j^{(2)}$ in Fig.~\ref{fig:Z2-lattice-w2}. As in the case of $\mathbb{Z}_2$ TI (see section~\ref{sec:QSH}), this implies that the decision boundary criterion for the $\mathbb{Z}_2$ lattice gauge theory
is also a non-trivial function of the inputs.

\begin{figure}
\includegraphics[scale=0.65]{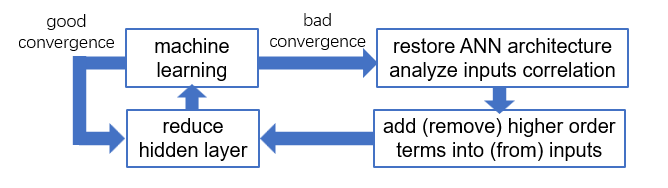}
\caption{A supervised machine learning framework that progressively includes higher order terms of the original inputs to handle the non-linearity, and reduces the hidden layer width for interpretability.}
\label{fig:non-linear-interpret}
\end{figure}

\begin{figure}
\includegraphics[scale=1.0]{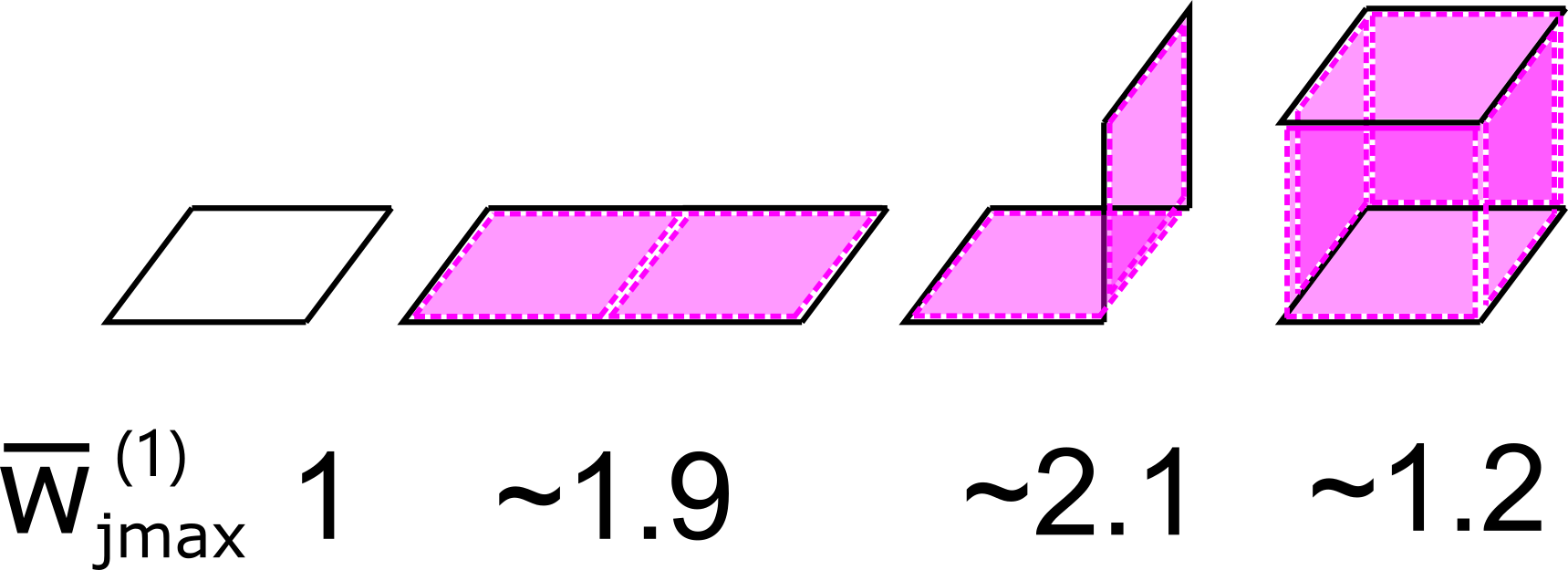}
\caption{Higher order terms of the original inputs showing strong correlations in the machine learning of the deconfinement of the $\mathbb Z_2$ lattice gauge theory are mostly the products of local plaquettes that give rise to larger Wilson loops. Such quadratic order terms are included as new inputs, progressively to preprocess the non-linearity, so that the complexity of the RELU ANN can be reduced for interpretation. The numbers below are the relative $w^{(1)}$ weights averaged over equivalent inputs under translations and rotations. }
\label{fig:larger-loops}
\end{figure}

To gain insight into the function that the ANN learns, we introduce non-linearity within the preprocessing so that the target function can be represented approximately linearly. For this, we include asymptotically higher-order terms of the normalized inputs $x_i$ as new inputs, in exchange for reducing the width of the hidden layer. This way, we aim to determine the nature of the non-linearity captured in Fig.~\ref{fig:Z2-lattice-w2}. In particular, we can include higher-order terms $x_i x_j$ to the input for further training steps when the inputs $x_i$ and $x_j$ show a strong correlation at the current step, for instance $y(x_i, x_j)+y(x_j, x_i)-2y(x_i/2+x_j/2, x_i/2+x_j/2)$ with all other inputs omitted in the expression. For simplicity, we limit ourselves to the quadratic order of the original inputs, related to $W_p$ for the study of the $\mathbb{Z}_2$ lattice gauge theory. In the meantime, we gradually reduce the hidden layer width of the RELU neural network to approach the linear limit, and eliminate newly-included inputs that do not contribute significantly to the output, reducing the width while maintaining performance, see Fig.~\ref{fig:non-linear-interpret}. When the supervised machine learning of the RELU neural network finally converges with a small hidden layer width, the ANN can be interpreted as in the linear function formalism of Sec.~\ref{sec:CI}. In comparison with Taylor expansion \cite{WangCe2019, Sebastian2017}, our anatomy of the ANN is both automated and relatively cost-efficient, as only higher-order contributions deemed relevant are retained in the end. Also, our algorithm mainly offers a pre-processing of the inputs, which can complement Taylor expansion of the ANN for a nonlinear decision function in terms of the inputs.

The above iterative approach singles out the products of neighboring inputs (see Fig.~\ref{fig:larger-loops}) as new inputs that simultaneously permit the hidden layer width to shrink to as narrow as 3 neuron-wide and contribute with the most weight. Here we averaged over the weights for the inputs of identical geometry to obtain the relative contributions $\bar w^{(1)}_{jmax}$ for each type of higher-order inputs. The selection of the higher-order inputs and their weights offer much insight into the learning of the ANN. Firstly, it is notable that products of more distant inputs are left out. The new higher-order inputs are exclusively those that combine two smaller loops from the initial input to form a larger loop. This is a feature that is commensurate with expectation for the Abelian gauge theory, for which small Wilson loops fuse to form larger Wilson loops. Secondly, the fact that the new larger loops acquire larger weight compared to the original small loop indicates that the ANN's criteria are consistent with expectation for a deconfinement transition of the $\mathbb{Z}_2$ lattice gauge theory. For a rigorous identification of the transition, the expectation value of the Wilson loop needs to be calculated for a large loop in the thermodynamic limit, a challenging task for any computational approach. But we have demonstrated that the ANN can discover the deconfinement transition by using small loops, together with slightly larger loops obtained through multiplication, to arrive at a rudimentary yet physically sound judgment.

\section{Conclusions}\label{sec:conclusion}

To summarize, we studied the interpretability of machine learning in the context of three distinct topological quantum phase transitions learned by shallow, fully connected feed-forward ANNs. The quantum phase transitions of interest are between topologically trivial states and a Chern insulator (CI), topological insulator (TI), and quantum spin liquid (QSL), respectively. As is well-known in the machine learning literature \cite{AppliedPM}, the more expressive the machine learning architectures are, the more opaque are their decision making criteria.
%To date, ``interpretable'' machine learning in problems of quantum matter has been restricted to linear models.
The relatively simple and minimally non-linear structure of our ANN was able to learn topological phases, aided by the quantum loop tomography (QLT): a physically motivated feature selection scheme. For the CI, the criteria the ANN learned amounted to evaluating a noisy version of the relevant topological invariant, which was linear in the QLT inputs. For the TI and the QSL, on the other hand, the ANNs based their criteria on non-linear functions of the inputs. For QSL, we determined that the ANN fused the neighboring Wilson loops of the corresponding $\mathbb{Z}_2$ gauge theory in the QLT to form larger Wilson loops.

Our successful interpretation of QLT-based machine learning gives us confidence in the ANN's phase detection by confirming that its decision criterion is guided by key physical properties of the target phases. Our results should serve to encourage wider application of QLT-based machine learning. We note the important role of physical insight that guided the design of QLT feature selection which, in turn, enabled interpretable machine learning. The shallow ANN depth, combined with physical insight, powered the QLT-enabled interpretation of the learning for the CI. Understanding of deconfinement in the thermodynamic limit was critical to the successful interpretation of fully non-linear machine learning of the QSL. Our approach to interpreting the learning of QSL bears similarity to the variational auto-encoder \cite{VAE1, VAE2}, and further investigating this similarity could be an interesting future direction. These results also provide hope that interpretable machine learning in the future can instead {\it inform\/} our physical insight, when our prior understanding is not sufficient to craft the necessary informative features. We could imagine instead such informative composite features emerging in later layers of a deep neural network fed only naive features, and whose interpretation would then lead to a better theoretical understanding of the underlying physics.

{\noindent \bf Acknowledgements} We thank Eduardo Fradkin for motivating this work. We thank P.E. Shanahan for useful discussions. Y.Z. and E.-A.K. acknowledge support by the U.S. Department of Energy, Office of Basic Energy Sciences, Division of Materials Science and Engineering under Award DE-SC0018946. Y.Z. also acknowledges support from the start-up grant at International Center for Quantum Materials and High-Performance Computing Platform of Peking University.

\bibliography{refs2}

\end{document}